\documentclass[12pt, a4paper]{article}

\usepackage[numbers]{natbib}
\usepackage{rotating}
\usepackage{graphicx}

\usepackage{ifthen} 
\newboolean{pdflatex}
\setboolean{pdflatex}{true} 

\newboolean{articletitles}
\setboolean{articletitles}{true} 

\newboolean{uprightparticles}
\setboolean{uprightparticles}{false} 

\newboolean{inbibliography}
\setboolean{inbibliography}{false} 

\usepackage{mciteplus}

\begin{document}

\title{Observation of Top Quark Production in the Forward Region at LHCb}

\graphicspath{{Figures/}}

\maketitle
\begin{center}
Stephen Farry$^{1}${\small \\On behalf of the LHCb collaboration \\ $^{1}$Department of Physics, University of Liverpool, L69 7ZE, United Kingdom.}
\end{center}

\begin{abstract}
Forward top production is observed, in the $\mu + b$ final state, with the  3 fb$^{-1}$ Run I dataset collected by the LHCb detector. The combined cross-section for $t\bar{t}$ and single top production at $\sqrt{s}$=7 TeV and $\sqrt{s}$=8 TeV is measured, for muons from the W boson with $p_{\rm T}>25$ GeV in the pseudo-rapidity range 2.0$<\eta<$4.5 and with a $b$-tagged jet with $50<p_{\rm T}<100$ GeV in the pseudorapidity range 2.2$<\eta<$4.2. The production cross-sections are found to be in agreement with NLO predictions.
\end{abstract}

\section{Introduction}
The LHCb detector~\cite{Alves:2008zz} is a dedicated forward detector at the LHC, fully instrumented in the pseudorapidity region $2.0<\eta<5.0$. It has been optimised to identify and reconstruct $b$ and $c$ hadron decays through precision tracking, vertexing and particle identification, and consequently is ideally suited to perform heavy flavour tagging of jets. The tagging of heavy flavour jets at LHCb, and specifically jets arising from $b$ quarks, can be used to identify and measure the production of top quarks in the forward region. Such a measurement, originally proposed in order to measure the asymmetry of $t\bar{t}$ production~\cite{PhysRevLett.107.082003}, has the potential to reduce the uncertainties on the gluon PDF by up to 20\% at large-$x$~\cite{Gauld:2013aja}. A study of the prospects for measuring top quark production at LHCb was performed in~\cite{Gauld:1557385}, with the number of events expected per fb$^{-1}$ of data shown in Table~\ref{tab:top_expevts}. It should be noted that the predictions are made for a specific fiducial region which differs from that used in the final analysis, and so the numbers act only as a guide. As the LHCb experiment collects a lower rate of luminosity than the ATLAS and CMS experiments, and has a smaller fiducial acceptance, a significantly lower number of top quark events are expected to be produced at LHCb during Run-I data taking. The most statistically accessible final state is that of a lepton and a $b$-jet, which also suffers from the lowest purity due to the large contribution from direct $Wb$ production. Nevertheless, this final state, and specifically that of a muon and a $b$-jet, is chosen to perform a measurement of top quark production in the forward rapidity region at LHCb~\cite{Aaij:2015mwa}, where top production includes contributions from both single top and $t\bar{t}$ production, with the latter contributing approximately 75\% of events. The heavy flavour tagging techniques used at LHCb are described first, followed by the selection criteria and purity extraction before finally the results obtained are discussed.
\begin{table}
\caption{Summary of the expected top production cross-sections at LHCb in different final states for centre-of-mass energies of 7, 8 and 14~TeV. The table is reproduced from Reference~\cite{Gauld:1557385}}
\label{tab:top_expevts}
\begin{center}
\begin{tabular}{c c c c}
$d\sigma$ (fb) & 7 TeV &  8 TeV & 14 TeV \\
\hline
$\ell b $ & 285 $\pm$ 52 & 504 $\pm$ 94 &  4366 $\pm$ 663 \\
$\ell bj$ & 97 $\pm$ 21 & 198 $\pm$ 35 & 2335 $\pm$ 323 \\
$\ell bb$ & 32 $\pm$ 6 & 65 $\pm$ 12 & 870 $\pm$ 116 \\
$\ell bbj$ & 10 $\pm$ 2 & 26 $\pm$ 4 & 487 $\pm$ 76 \\
$\ell\ell$ & 44 $\pm$ 9 & 79 $\pm$ 15 & 635 $\pm$ 109 \\
$\ell\ell b$ & 19 $\pm$ 4 & 39 $\pm$ 8 & 417 $\pm$ 79
\end{tabular}
\end{center}
\end{table}

\section{Heavy Flavour Tagging}
\begin{figure}
\includegraphics[width=0.32\textwidth]{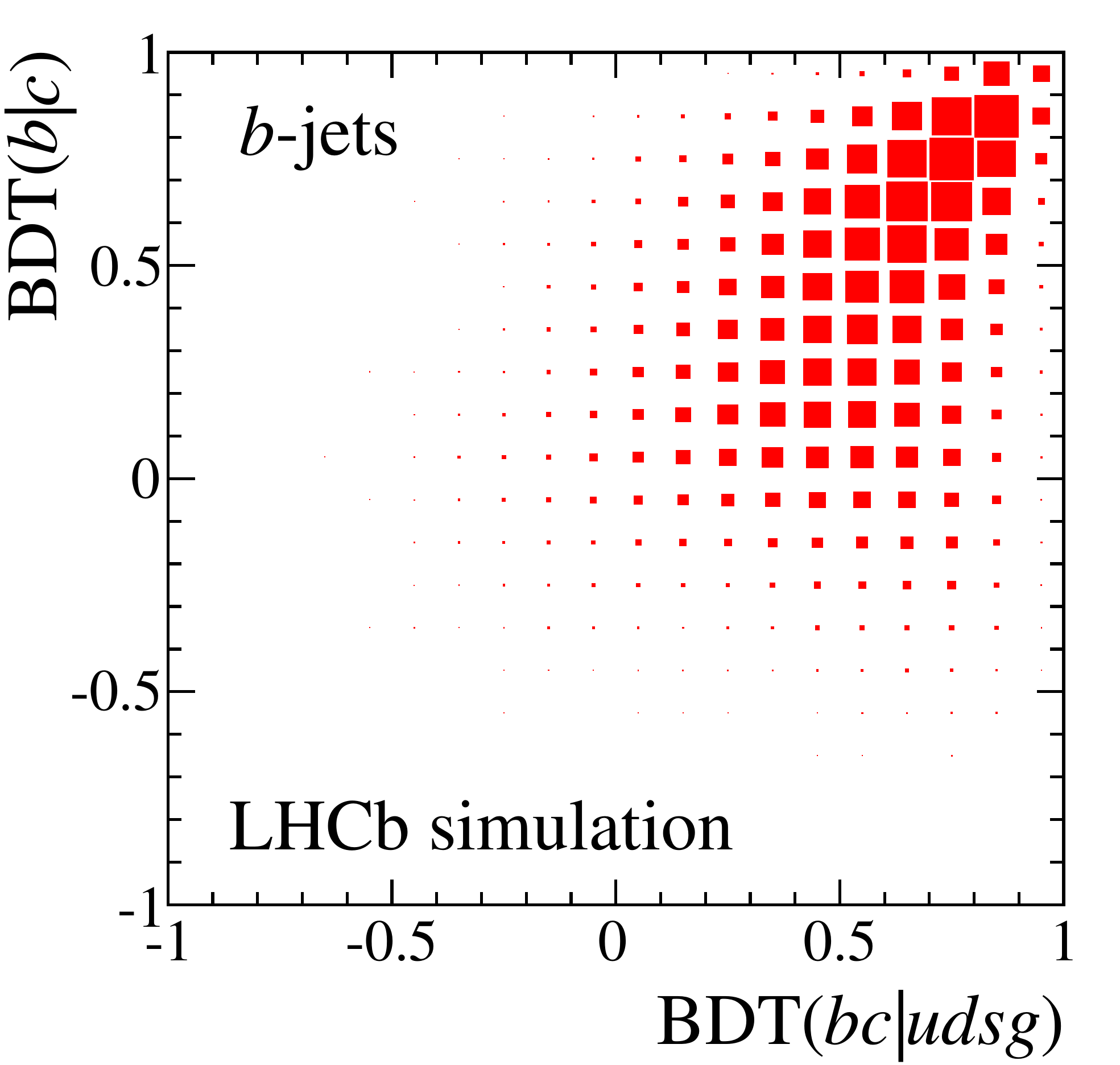}
\includegraphics[width=0.32\textwidth]{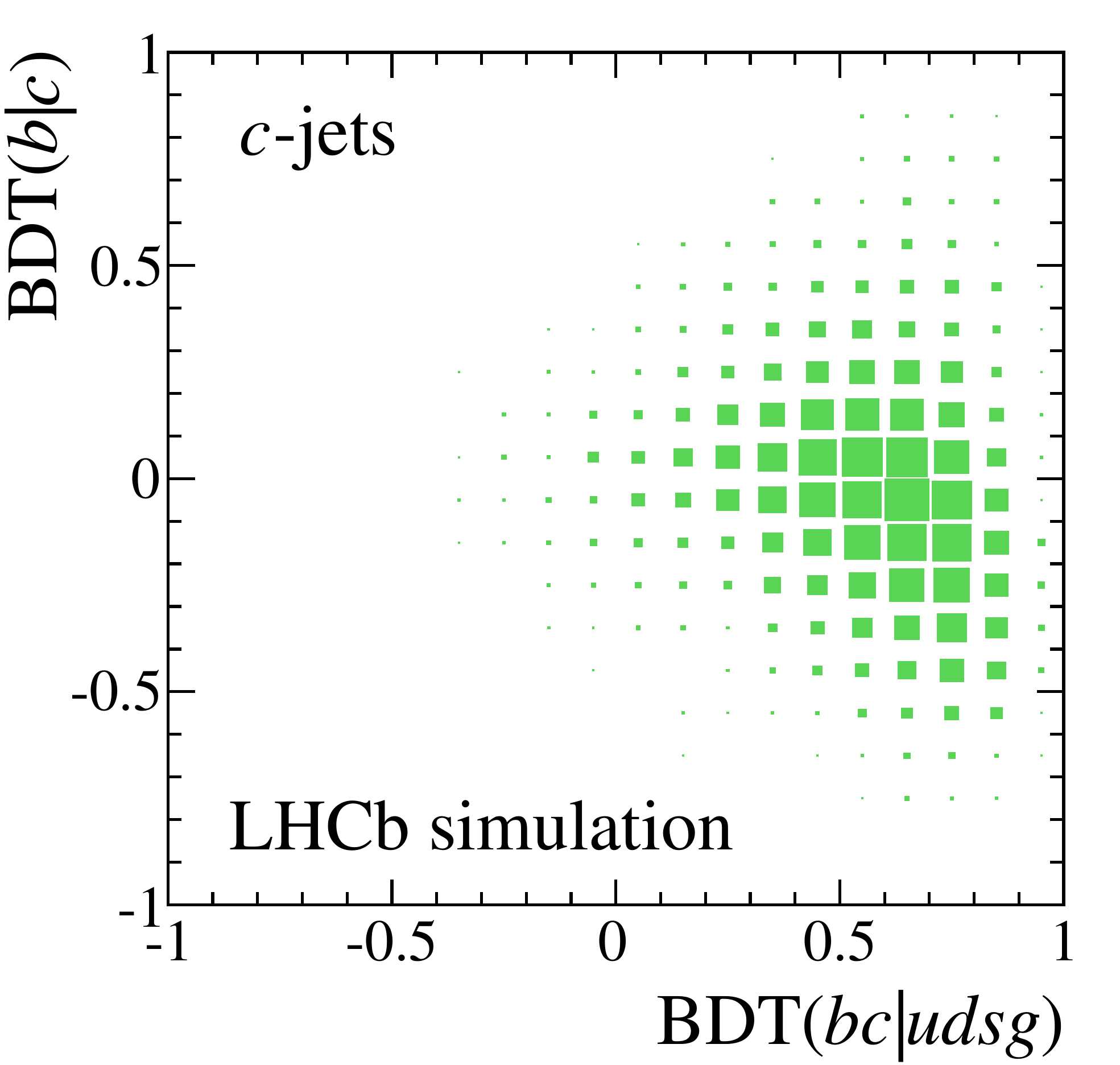}
\includegraphics[width=0.32\textwidth]{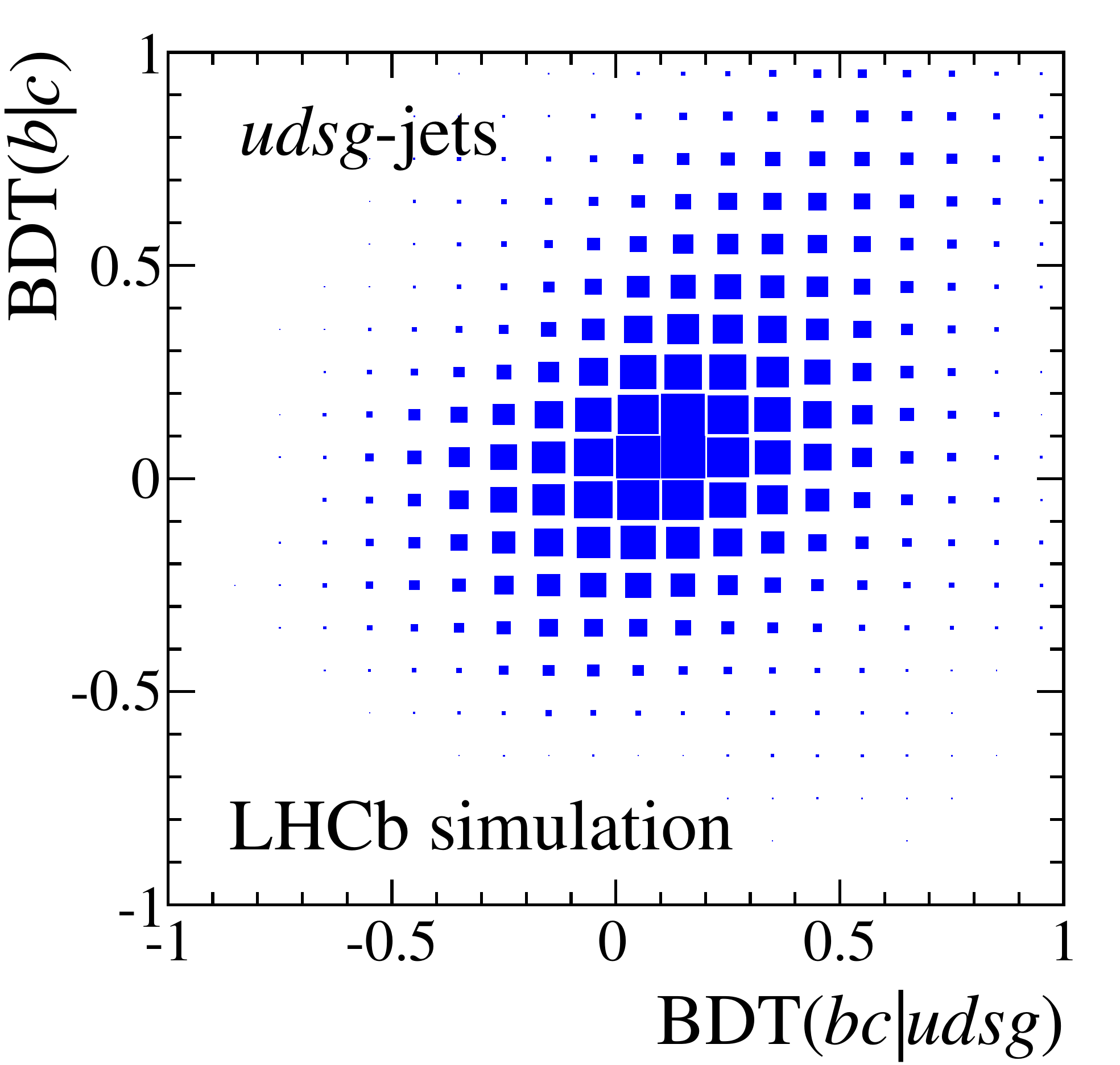}
\caption{A two-dimensional representation of the BDT responses in simulation is shown for (left) $b$-jets, (middle) $c$-jets and (right) light-jets.}
\label{fig:svtag}
\end{figure}

\begin{figure}
\begin{center}
\includegraphics[width=0.7\textwidth]{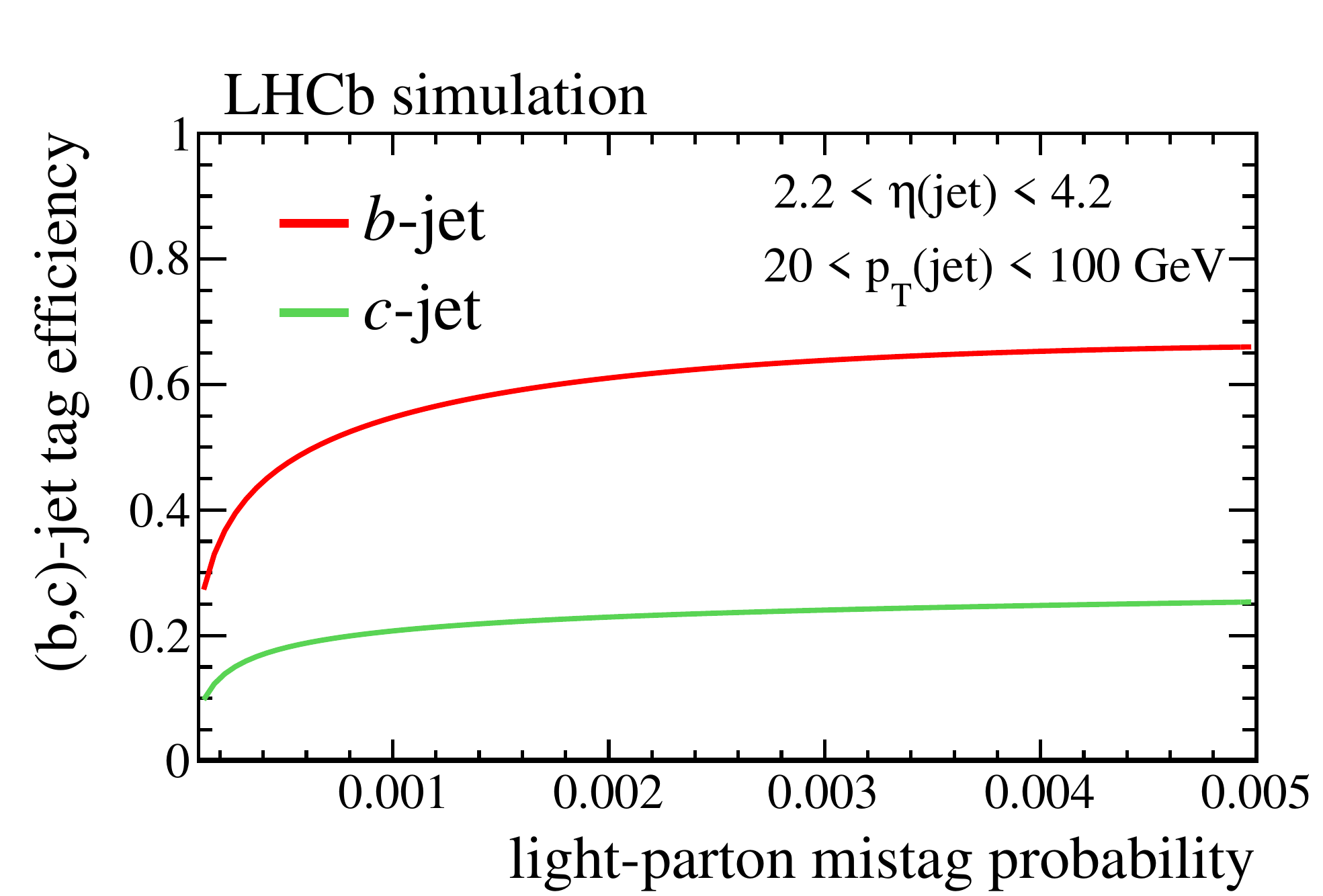}
\end{center}
\caption{The efficiency for tagging $b$ and $c-$jets versus the mis-tag probability for light jets obtained from simulation.}
\label{fig:bdteff}
\end{figure}

Heavy flavour tagging at LHCb is performed using the secondary vertex tagging algorithm outlined in~\cite{Aaij:2015yqa}. Jets are tagged first by searching for a secondary vertex (SV) within the jet which satisfies specific track and vertex quality requirements. Two boosted decision trees (BDTs) are then trained on simulated $b$, $c$, and light jet samples using properties of the SV and the jet in order to separate light jets from heavy quark jets ( BDT($bc|udsg$) ), and $b-$jets from $c-$jets ( BDT($b|c$) ). The primary variables used are related to the $b-$ or $c-$hadron decay as these are expected to be well modelled in simulation. One parameter used to discriminate the jet types is the corrected mass, $M_{\rm cor}$, defined as
$$
M_{\rm cor} = \sqrt{M^2 + p^2\sin^2\theta} + p\sin\theta
$$
where $M$ and $p$ are the mass and momentum of the particles that form the SV and $\theta$ is the angle between the flight direction of the vertex and its momentum. It represents the minimum mass the long-lived object decaying at the vertex can have which is consistent with the flight direction. Other parameters include the fraction of the jet $p_{\rm T}$ carried by the particles forming the SV, as well as its multiplicity, net charge and flight distance. A two-dimensional representation of the BDT responses for $b$, $c$ and light jets is shown in Figure~\ref{fig:svtag}. The output generated by the BDTs can then be used either to place requirements on the jets, or to perform a template fit to extract the relative amounts of $b$, $c$ and light jets in a selected sample. Where requirements are placed on the BDT($bc|udsg$) response, the tagging efficiency versus light jet mis-tag rate is shown for $b$ and $c-$jets in Figure~\ref{fig:bdteff}.

\section{Dataset and Selection}
The measurement is performed using data collected at centre-of-mass energies of 7 and 8 TeV, corresponding to integrated luminosities of approximately 1 and 2 fb$^{-1}$ respectively. The selection is based on previous studies of $W$ production in association with $b$ and $c-$jets performed at LHCb using the same dataset~\cite{Aaij:2015cha}. Events are selected which contain a muon with a transverse momentum, $p_{\rm T}(\mu)$, of greater than 25~GeV in the pseudorapidity region $2<\eta<4.5$, in addition to a jet satisfying $50 <p_{\rm T}(j)<100$~GeV in the pseudorapidity range $2.2<\eta<4.2$. The inputs for jet reconstruction are selected using a particle flow algorithm as described in ~\cite{Aaij:2013nxa}and jet are clustered using the anti-$k_{\rm T}$ algorithm with distance parameter, $R$ = 0.5. The muon and the jet are also required to be ``unbalanced'' in $p_{\rm T}$ by requiring that $p_{\rm T}(j_\mu + j)$, representing the $p_{\rm T}$ of the vectorial sum of $j_\mu$ and $j$ is greater than 20~GeV, where $j_\mu$ is a reconstructed jet containing the muon candidate and $j$ is the associated jet. The observable is expected to be large for $W+$jet events due to the missing neutrino in the final state and small for QCD di-jet production where the jet momenta are balanced. This represents a tighter fiducial selection than that used previously by LHCb to measure direct $Wb$ production, and is specifically chosen to enhance the expected contribution from top quark production.

\section{Purity Determination}
\begin{figure}
\includegraphics[width=0.5\textwidth]{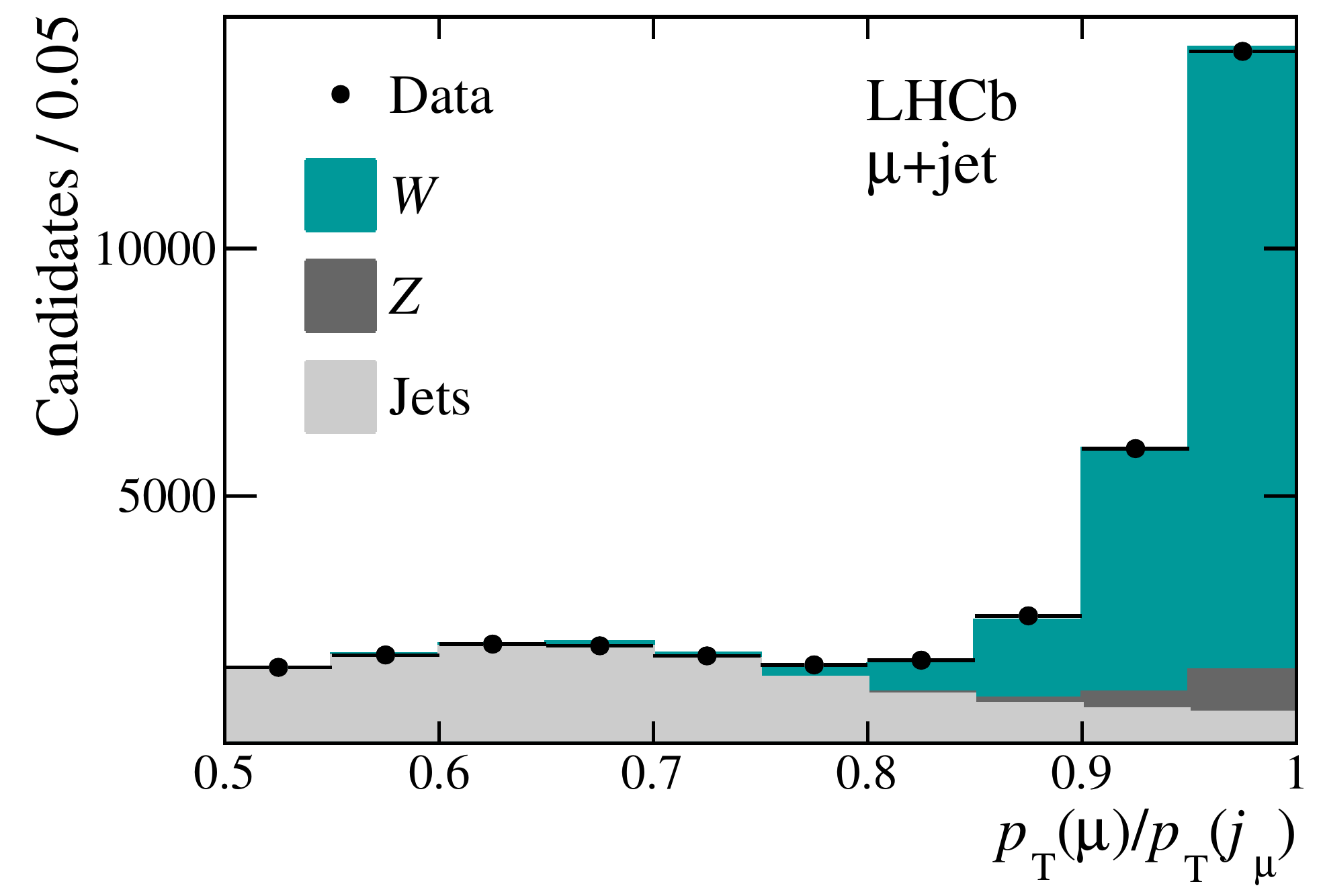}
\includegraphics[width=0.5\textwidth]{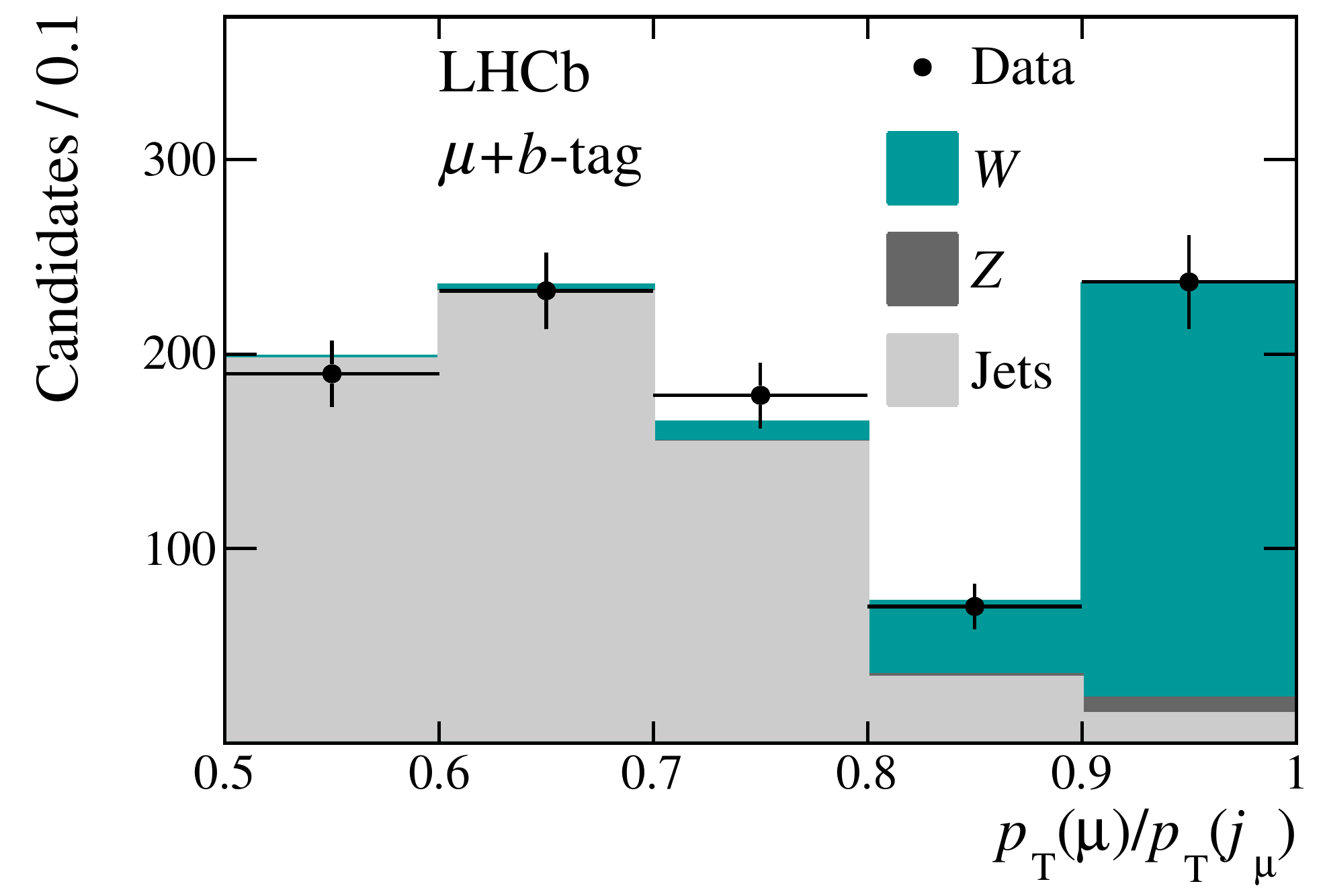}
\caption{The combined fits used to extract the sample purity for (left) the total $Wj$ sample and (right) the extracted $Wb$ component.}
\label{fig:purity}
\end{figure}
The purity is determined using a template fit to the isolation variable $p_{\rm T}(\mu)/p_{\rm T}(j_\mu)$, where $p_{\rm T}(\mu)$ is the transverse momentum of the muon in the final state, and $p_{\rm T}(j_\mu)$ is the transverse momentum of $j_{\mu}$. This variable is expected to peak towards unity for signal events, and to be spread to lower values for backgrounds arising from QCD multi-jet processes. The electroweak template shapes are taken from simulation and corrected for differences between data and simulation using $Z\to\mu\mu$ events. The QCD background is estimated using a data-driven method. Events are selected in a sideband region obtained by requiring the muon and jet momenta are balanced ($p_{\rm T}(j_\mu + j) < $20~GeV). This region is dominated by QCD events, and the events selected in this region are then reweighted in $p_{\rm T}(j_\mu)$ to match the shape obtained in the isolated signal region. In each bin of $p_{\rm T}(\mu)/p_{\rm T}(j_\mu)$, the contribution from $b$-jets is extracted by requiring that the jets contain an SV and performing a template fit to the resultant two-dimensional BDT distributions. The fits are performed separately for positive and negative muons, and for the different centre-of-mass energies. The combined fits are shown in Figure~\ref{fig:purity} for the full sample and the extracted $b$-jet component.

\section{Results}
\begin{figure}
\begin{center}
\includegraphics[width=0.6\textwidth]{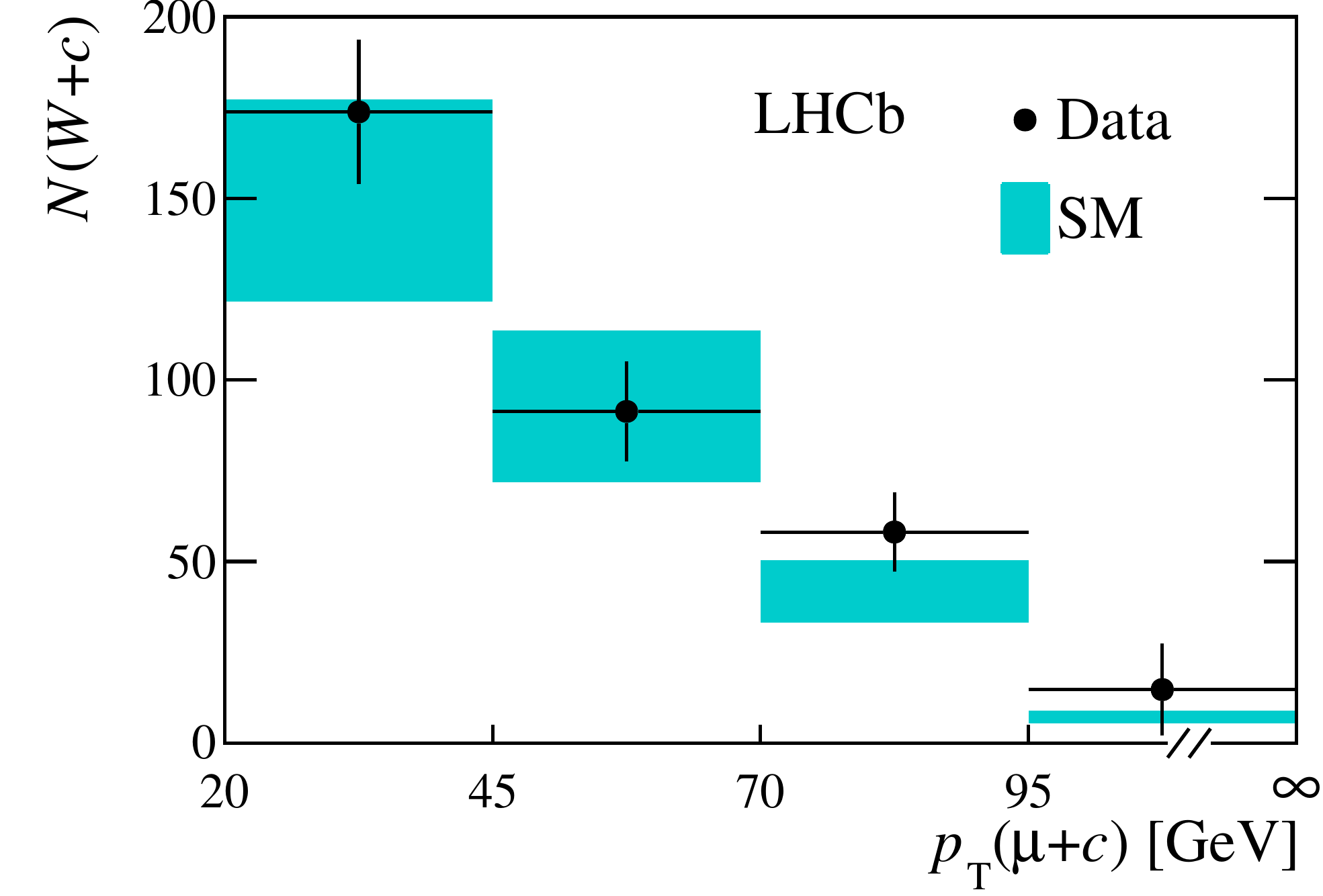}
\end{center}
\caption{The observed $Wc$ event yield compared with SM predictions obtained by first measuring $Wj$ production and scaling by the SM expectation for $\sigma(Wc)/\sigma(Wj)$ and the expected $c-$tagging efficiency.}
\label{fig:wc}
\end{figure}

\begin{figure}
\includegraphics[width=0.5\textwidth]{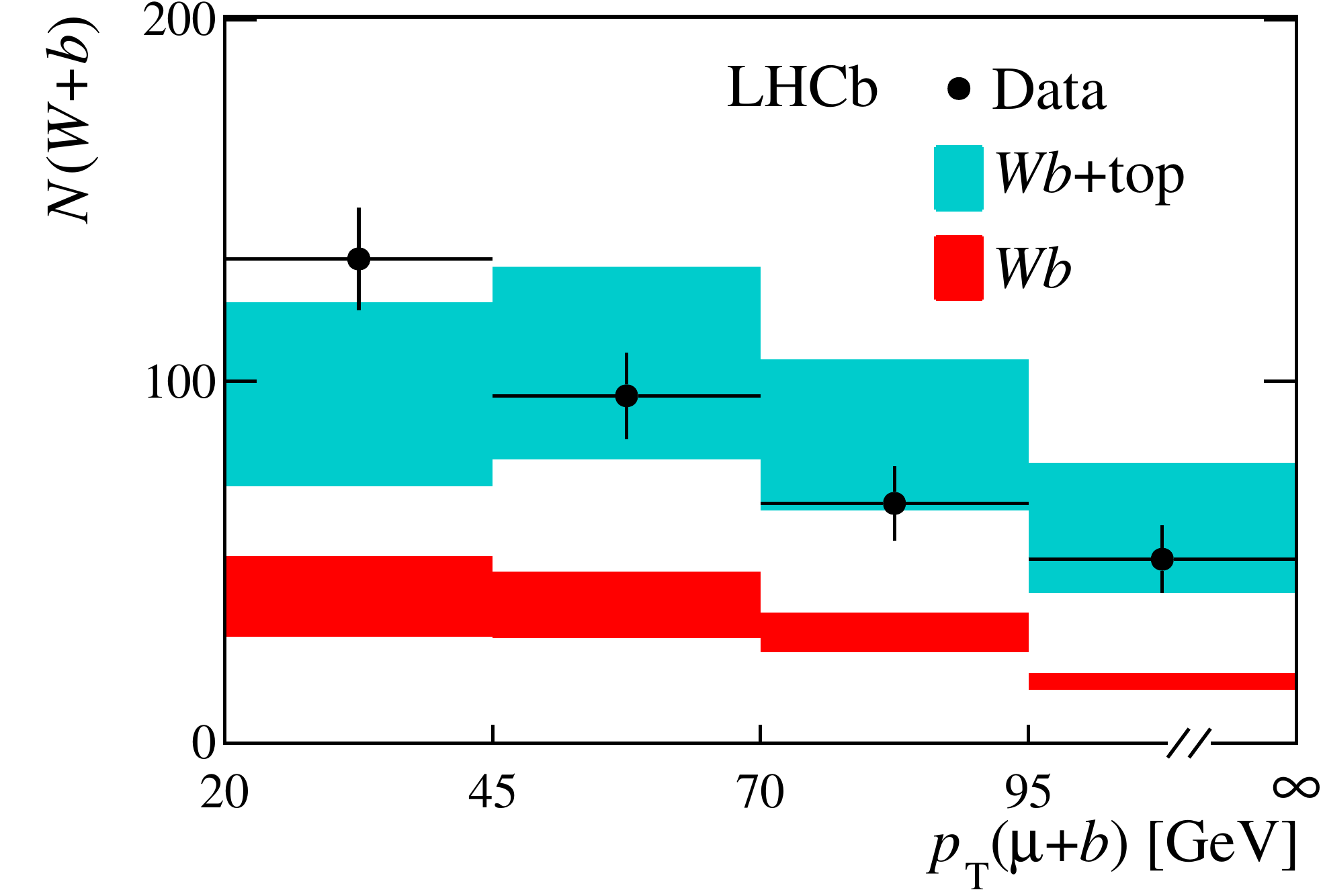}
\includegraphics[width=0.5\textwidth]{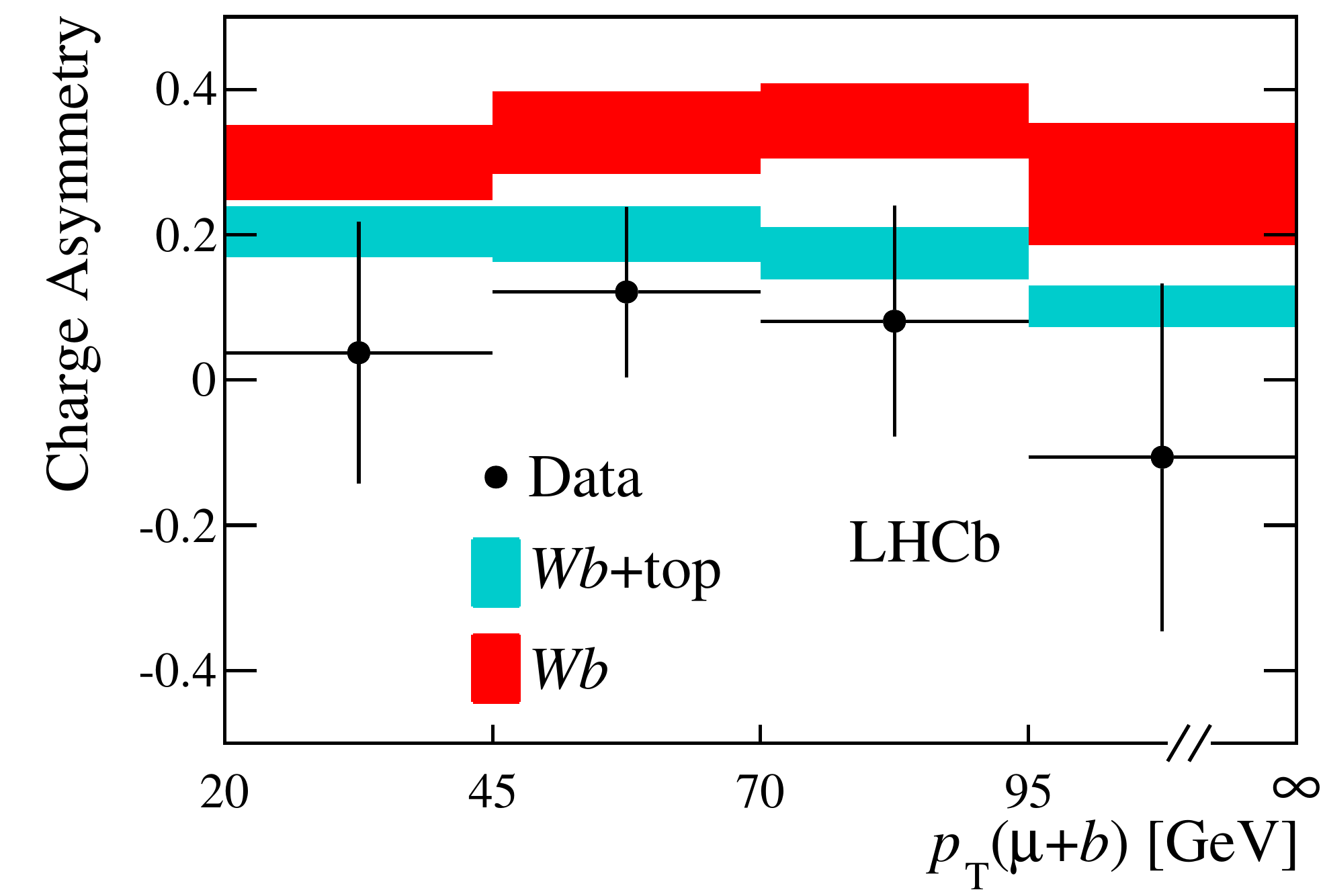}
\caption{The observed (left) event yield and (right) charge asymmetry of $Wb$ production compared to SM predictions calculated with and without the contribution from top quark production.}
\label{fig:results}
\end{figure}

The significance of the top quark contribution to the selected data sample is determined by comparing the observed event yield and charge asymmetry of $Wb$ production to the SM prediction with and without the contribution from top quark production. The predictions are calculated at NLO in perturbative QCD using the MCFM generator and folded for detector effects. As predictions for the ratio of $Wb$ to $Wj$ production are approximately a factor of three more precise than for $Wb$ production alone, the expected contribution from direct $Wb$ production is obtained by first measuring the $Wj$ cross-section in the chosen fiducial region, and then scaling it by the SM prediction for $\sigma(Wb)/\sigma(Wj)$ and the expected $b-$tagging efficiency. This approach is verified using the $Wc$ channel, where no extra backgrounds are expected and good agreement is seen between the prediction obtained and the measurement, as shown in Figure~\ref{fig:wc}. The observed event yield and charge asymmetry are shown in Figure~\ref{fig:results} where a factor of three more events are observed than would be expected for the SM excluding top production, as well as a lower charge asymmetry.   The statistical significance of the top quark contribution is calculated using a binned profile likelihood test to compare the data to the SM hypothesis without a top quark contribution and where contributions from both single top and $t\bar{t}$ production are included. A number of sources of systematic uncertainty are considered and are tabulated in Table~\ref{tab:sys}, with the uncertainty on the $b-$tagging efficiency seen to dominate. The significance obtained is 5.4$\sigma$, confirming the observation of top quark production in the forward region. The excess is then used to calculate the cross-section for top quark production in the chosen fiducial region, which mirrors the kinematic selection, except for the requirement on $p_{\rm T}(j_\mu + j)$, which is alternatively applied to $p_{\rm T}(\mu + j)$. The cross-section includes contributions from both $t\bar{t}$ and single top production, and is determined to be
\begin{eqnarray*}
\sigma(\mathrm{top})[7\mathrm{ TeV}] &=& 239 \pm 53 \mathrm{(stat)} \pm 33 \mathrm{(syst)} \pm 24 \mathrm{(theory)} \mathrm{fb} \\
\sigma(\mathrm{top})[8\mathrm{ TeV}] &=& 289 \pm 43 \mathrm{(stat)} \pm 40 \mathrm{(syst)} \pm 29 \mathrm{(theory)} \mathrm{fb,}
\end{eqnarray*}
where the first uncertainty is statistical, the second is systematic, and the third is due to uncertainties in the theoretical modelling. The results are in agreement with the SM predictions.

\begin{table}
    \caption{\label{tab:sys} Relative systematic uncertainties. The symbol $\dagger$ denotes an uncertainty that only applies to the cross-section measurement and not the significance determination.}
    \newcommand{\pad}{\phantom{$^\dagger$}}
    \begin{center}
    \begin{tabular}{lr}
      source & uncertainty \\
	\hline
      GEC &  2\%\pad \\
      $p_{\rm T}(\mu)/p_{\rm T}(j_\mu)$ templates & 5\%\pad \\
      jet reconstruction & 2\%\pad \\
      SV-tag BDT templates & 5\%\pad \\
      $b$-tag efficiency & 10\%\pad \\
      trigger \& $\mu$ selection & $2\%^{\dagger}$ \\
      jet energy & $5\%^{\dagger}$ \\
      $W\to\tau\to\mu$ & $1\%^{\dagger}$ \\
      luminosity & 1--2\%$^{\dagger}$ \\
	\hline
      Total & 14\% \\
      \hline
      Theory & 10\% \\
    \end{tabular}
    \end{center}
\end{table}

\section{Conclusion}
The first observation of top quark production is performed in the forward region using the Run-I dataset collected at LHCb. The measurement is performed in the $\mu + b-$jet final state and is found to be in good agreement with SM predictions. While measurements are currently statistically limited, a large increase in the top quark production cross-section is expected in Run-II at the higher centre-of-mass energy of 13~TeV. Consequently, a number of different final states will become statistically accessible and will allow a higher purity to be achieved as well as the separation of $t\bar{t}$ and single top production.



\bibliographystyle{LHCb}%
\bibliography{lhcp}%

\ifx\mcitethebibliography\mciteundefinedmacro
\PackageError{LHCb.bst}{mciteplus.sty has not been loaded}
{This bibstyle requires the use of the mciteplus package.}\fi
\providecommand{\href}[2]{#2}
\begin{mcitethebibliography}{1}
\mciteSetBstSublistMode{n}
\mciteSetBstMaxWidthForm{subitem}{\alph{mcitesubitemcount})}
\mciteSetBstSublistLabelBeginEnd{\mcitemaxwidthsubitemform\space}
{\relax}{\relax}

\bibitem{Alves:2008zz}
LHCb, A.~A. Alves, Jr.\ {\em et~al.},
  \ifthenelse{\boolean{articletitles}}{\emph{{The LHCb Detector at the LHC}},
  }{}\href{http://dx.doi.org/10.1088/1748-0221/3/08/S08005}{JINST \textbf{3}
  (2008) S08005}\relax
\mciteBstWouldAddEndPuncttrue
\mciteSetBstMidEndSepPunct{\mcitedefaultmidpunct}
{\mcitedefaultendpunct}{\mcitedefaultseppunct}\relax
\EndOfBibitem
\bibitem{PhysRevLett.107.082003}
A.~L. Kagan, J.~F. Kamenik, G.~Perez, and S.~Stone,
  \ifthenelse{\boolean{articletitles}}{\emph{Probing new top physics at the
  lhcb experiment},
  }{}\href{http://dx.doi.org/10.1103/PhysRevLett.107.082003}{Phys.\ Rev.\
  Lett.\  \textbf{107} (2011) 082003}\relax
\mciteBstWouldAddEndPuncttrue
\mciteSetBstMidEndSepPunct{\mcitedefaultmidpunct}
{\mcitedefaultendpunct}{\mcitedefaultseppunct}\relax
\EndOfBibitem
\bibitem{Gauld:2013aja}
R.~Gauld, \ifthenelse{\boolean{articletitles}}{\emph{{Feasibility of top quark
  measurements at LHCb and constraints on the large-$x$ gluon PDF}},
  }{}\href{http://dx.doi.org/10.1007/JHEP02(2014)126}{JHEP \textbf{02} (2014)
  126}, \href{http://arxiv.org/abs/1311.1810}{{\normalfont\ttfamily
  arXiv:1311.1810}}\relax
\mciteBstWouldAddEndPuncttrue
\mciteSetBstMidEndSepPunct{\mcitedefaultmidpunct}
{\mcitedefaultendpunct}{\mcitedefaultseppunct}\relax
\EndOfBibitem
\bibitem{Gauld:1557385}
R.~Gauld, \ifthenelse{\boolean{articletitles}}{\emph{{Measuring top quark
  production asymmetries at LHCb}}, }{} Tech. Rep. LHCb-PUB-2013-009.
  CERN-LHCb-PUB-2013-009, CERN, Geneva, Jun, 2013\relax
\mciteBstWouldAddEndPuncttrue
\mciteSetBstMidEndSepPunct{\mcitedefaultmidpunct}
{\mcitedefaultendpunct}{\mcitedefaultseppunct}\relax
\EndOfBibitem
\bibitem{Aaij:2015mwa}
LHCb, R.~Aaij {\em et~al.}, \ifthenelse{\boolean{articletitles}}{\emph{{First
  observation of top quark production in the forward region}},
  }{}\href{http://arxiv.org/abs/1506.00903}{{\normalfont\ttfamily
  arXiv:1506.00903}}\relax
\mciteBstWouldAddEndPuncttrue
\mciteSetBstMidEndSepPunct{\mcitedefaultmidpunct}
{\mcitedefaultendpunct}{\mcitedefaultseppunct}\relax
\EndOfBibitem
\bibitem{Aaij:2015yqa}
LHCb, R.~Aaij {\em et~al.},
  \ifthenelse{\boolean{articletitles}}{\emph{{Identification of beauty and
  charm quark jets at LHCb}},
  }{}\href{http://dx.doi.org/10.1088/1748-0221/10/06/P06013}{JINST \textbf{10}
  (2015), no.~06 P06013},
  \href{http://arxiv.org/abs/1504.07670}{{\normalfont\ttfamily
  arXiv:1504.07670}}\relax
\mciteBstWouldAddEndPuncttrue
\mciteSetBstMidEndSepPunct{\mcitedefaultmidpunct}
{\mcitedefaultendpunct}{\mcitedefaultseppunct}\relax
\EndOfBibitem
\bibitem{Aaij:2015cha}
LHCb, R.~Aaij {\em et~al.}, \ifthenelse{\boolean{articletitles}}{\emph{{Study
  of $W$ boson production in association with beauty and charm}},
  }{}\href{http://arxiv.org/abs/1505.04051}{{\normalfont\ttfamily
  arXiv:1505.04051}}\relax
\mciteBstWouldAddEndPuncttrue
\mciteSetBstMidEndSepPunct{\mcitedefaultmidpunct}
{\mcitedefaultendpunct}{\mcitedefaultseppunct}\relax
\EndOfBibitem
\bibitem{Aaij:2013nxa}
LHCb, R.~Aaij {\em et~al.}, \ifthenelse{\boolean{articletitles}}{\emph{{Study
  of forward Z + jet production in pp collisions at $\sqrt{s} = 7$ TeV}},
  }{}\href{http://dx.doi.org/10.1007/JHEP01(2014)033}{JHEP \textbf{01} (2014)
  033}, \href{http://arxiv.org/abs/1310.8197}{{\normalfont\ttfamily
  arXiv:1310.8197}}\relax
\mciteBstWouldAddEndPuncttrue
\mciteSetBstMidEndSepPunct{\mcitedefaultmidpunct}
{\mcitedefaultendpunct}{\mcitedefaultseppunct}\relax
\EndOfBibitem
\end{mcitethebibliography}

\end{document}